\documentclass[prl,aps, twocolumn, showpacs]{revtex4-1}

\usepackage{latexsym} \usepackage{graphicx}
\usepackage{amssymb} \usepackage{bm}
\usepackage{color}\usepackage{amsmath}
\newcommand{\blue}[1]{\textcolor[rgb]{0.00,0.00,1.00}{#1}}

\begin{document}

\title{Clustering of Magnetic Swimmers in a Poiseuille Flow}
\author{Fanlong Meng}
\affiliation{Rudolf Peierls Centre for Theoretical Physics, University of Oxford, Oxford OX1 3NP, UK}
\author{Daiki Matsunaga}
\affiliation{Rudolf Peierls Centre for Theoretical Physics, University of Oxford, Oxford OX1 3NP, UK}
\author{Ramin Golestanian}
\email[]{ramin.golestanian@physics.ox.ac.uk}
\affiliation{Rudolf Peierls Centre for Theoretical Physics, University of Oxford, Oxford OX1 3NP, UK}

\date{\today}

\begin{abstract}
We investigate the collective behavior of magnetic swimmers, which are suspended in a Poiseuille flow and placed under an external magnetic field, using analytical techniques and Brownian dynamics simulations. We find that the interplay between intrinsic activity, external alignment, and magnetic dipole-dipole interactions leads to %a rich variety of dynamical regimes with various forms of radial and
longitudinal structure formation. Our work sheds light on a recent experimental observation of a clustering instability in this system.
%\blue{We argue that hydrodynamic interactions play a secondary role in this between magnetic swimmers influence cluster formation.}
\end{abstract}

\maketitle

The intrinsic nonequilibrium activity of microscopic swimmers and the many possible means of interaction between them have provided us with an exquisite playground for studying emergent properties of active matter~\cite{Ramaswamy2010,Marchetti2013}. Examples of such systems include those with short range interactions and activity mechanisms that exhibit polar \cite{Vicsek1995,Toner1998} and nematic \cite{Chate2006, Bertin2013,Zhou2014,Thampi2016,Guillamat2016,Genkin2017} symmetry, as well as those with intrinsic mechanisms that mediate nonequilibrium long-range interactions such as phoretic swimmers \cite{Golestanian2012,Saha2014,Cohen2014}. %active gels~\cite{Kohler2011,Prost2015}
On the other hand, external cues can guide individual microswimmers in nature by using different strategies to modulate their motility patterns, including chemotaxis~\cite{Darnton2007,Cluzel2004,Lambert2010,Gachelin2014}, phototaxis \cite{Garcia2013,Jibuti2014,Lauga2017}, gyrotaxis \cite{Kessler1986,Pedley1990,Croze2017}, and magnetotaxis~\cite{Waisbord2016}. In the quest for understanding the full range of emergent properties of active matter systems, it is important to study the interplay between intrinsic mechanical activity, interactions,
and external driving, including shear stresses \cite{Zottl2012}. Magnetic active matter constitutes a particularly interesting class of systems, not least because of its versatility in terms of the fabrication of artificial swimmers and their control, as well as its potential for technological applications~\cite{Dreyfus2005,Tierno2008,Ogrin2008,Namdeo2013,Walker2015,Babel2016,Guzman-Lastra2016,Chen2017}.

\begin{figure}[b]%[h]
\begin{center}
\includegraphics[width=0.85\columnwidth]{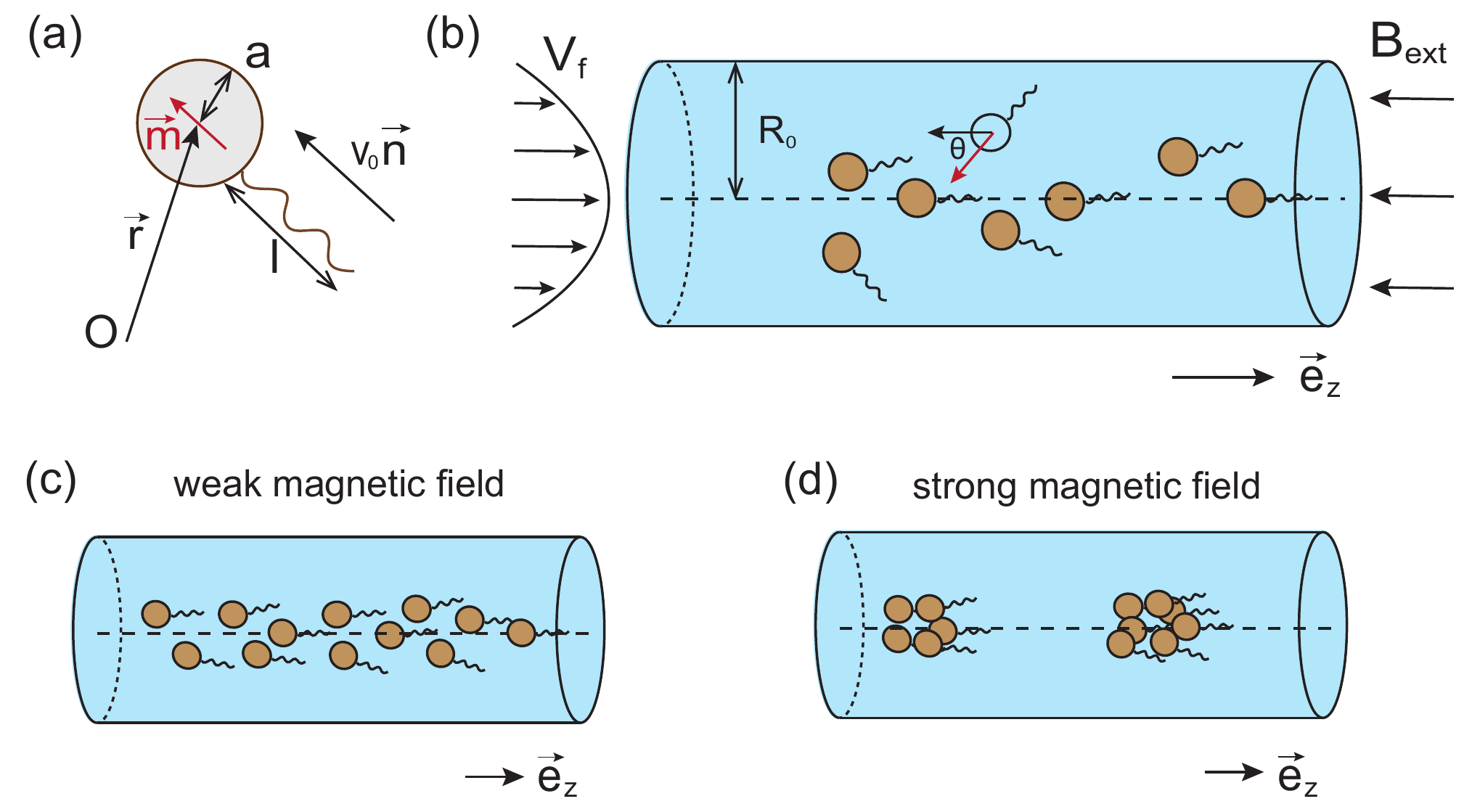}
\caption{(a) A magnetic swimmer consisting of a spherical head of radius $a$ and a helical tail of length $l$, which is located at $\bm{r}$ and pointing along $\bm{n}$. The swimming velocity of the swimmer is $v_{0}\bm{n}$ and the magnetic moment is $m_{0}\bm{n}$. (b) Magnetic swimmers in a Poiseuille flow (with velocity profile $\bm{V}_{\mathrm{f}}$) under an external magnetic field, confined in a circular channel of radius $R_{0}$ and length $L$. $\theta$ denotes the angle between the external magnetic field and the magnetic moment. (c) Focusing and (d) clustering of magnetic swimmers under weak and strong magnetic fields, respectively. }
\label{sketch}
\end{center}
\end{figure}

In driven active suspensions through channels, such as phototactic algae exposed to a light source~\cite{Garcia2013,Jibuti2014} and magnetotactic bacteria under the effect of an external magnetic field~\cite{Waisbord2016}, the swimmers have been observed to accumulate at the center and exhibit clustering. While the clustering can be understood as a consequence of effectively attractive hydrodynamic interactions between pullers such as algae~\cite{Lauga2017}, for the magnetotactic bacteria studied in Ref. \cite{Waisbord2016}, which have the structural characteristics of pusher-type bacteria such as \emph{E. coli}, the effect cannot be accounted for using hydrodynamic interactions. Here, we study the behavior of interacting magnetic swimmers under the effect of an external magnetic field and external shear flow; see Fig. \ref{sketch}. We show that the system, which has similarities to the class of phoretic swimmers \cite{Golestanian2012,Saha2014,Cohen2014}, exhibits a rich phenomenology, and provide an explanation for the mechanism behind the experimentally observed clustering instability reported in Ref. \cite{Waisbord2016}.

\paragraph{Description of the model.---}
We consider a self-propelled magnetic swimmer, which consists of a spherical head of radius $a$ that contains a permanent magnetic moment $m_{0}$ pointing in the direction of its body axis $\bm{n}$ as a simplified model of the magnetosome~\cite{Blakemore1975,Faivre2008,Klumpp2016}, and a helical flagellum of length $l$, as sketched in Fig.~\ref{sketch}(a). For simplicity, we ignore the anisotropy of the body shape and use scalar translational and rotational friction coefficients for the swimmer, denoted as $\zeta$ and $\zeta_{r}$, respectively (in general they depend on $a$ and $l$).
Also, we assume that the magnetic dipole is point-like and located at the center of the swimmer head. Furthermore, we assume that $\bm{n}$ is also the direction of propulsion of the swimmer, with the swimming speed of $v_{0}$. If there is an external magnetic field, $\bm{B}_{\mathrm{ext}}$, the magnetic moment in the swimmer will tend to align with it, and thus the swimmer will move in the same direction as the magnetic field; i.e. the swimmer is magnetotactic.

Suppose that in a cylindrical channel of radius $R_{0}$ and length $L$, there is a Poiseuille flow along the longitudinal direction of the channel $\bm{e}_{z}$, which together with $\bm{e}_{r}$ and $\bm{e}_{\phi}$ (the unit vectors along the radial and the azimuthal direction of the channel, respectively) form a complete basis set describing a cylindrical coordinate system. The velocity profile can be described as a function of the radial position in the channel as $\bm{V}_{\mathrm{f}}(r)=\bm{e}_{z} v_{\mathrm{f}}(R_{0}^{2}-r^{2})/R_{0}^{2}$, where $v_{\mathrm{f}}$ is the maximum flow speed at the center of the channel [Fig.~\ref{sketch}(b)].
If the magnetic swimmers are suspended in the above flow, and there is an external magnetic field  $\bm{B}_{\mathrm{ext}}$ (applied in $-\bm{e}_{z}$ direction), then the time evolution of the position $\bm{r}$ and the orientation $\bm{n}$ of a magnetic swimmer will be governed by the following equations
\begin{eqnarray}
\frac{d\bm{r}}{dt}&=&v_{0}\bm{n}+\bm{V}_{\mathrm{f}}+\frac{1}{\zeta} \bm{\nabla}\big(m_{0}\bm{n}\cdot \bm{B}_{\mathrm{int}}\big) + \bm{\xi},\label{L1}\hskip2.2cm\\
\frac{d\bm{n}}{dt}&=&\left[\frac{m_{0}}{\zeta_{r}}\bm{n}\times(\bm{B}_{\mathrm{ext}}+\bm{B}_{\mathrm{int}})+\frac{1}{2} \bm{\nabla} \times \bm{V}_{\mathrm{f}}+ \bm{\xi}_{r}\right]\times\bm{n},\label{L2}
\end{eqnarray}
where $\bm{\xi}$ and $\bm{\xi}_{r}$ denote translational and rotational Gaussian white thermal noise terms, respectively, with their strengths controlled by the corresponding diffusion coefficients $D=k_{\rm B} T/\zeta$ and $D_r=k_{\rm B} T/\zeta_r$. %,\emph{i.e.}, $-m_{0}\bm{n}\cdot \bm{B}_{\mathrm{int}}$ denotes magnetic dipole-dipole interactive energy between magnetic swimmers.
The internal magnetic field, $\bm{B}_{\mathrm{int}}$, which is induced by the other swimmers in the suspension, can be obtained from Amp\`ere's law
%as $\bm{B}_{\mathrm{int}}=\bm{\nabla}\times\bm{A}_{\mathrm{int}}$, where the magnetic vector potential $\bm{A}_{\mathrm{int}}$ obeys
\cite{Jackson1998}
\begin{equation}\label{bmtor_potential}
\bm{\nabla}\times\bm{B}_{\mathrm{int}}=\mu_{0}\bm{\nabla}\times\int d\bm{n}\; (m_{0}\bm{n}) \; P(\bm{r}, \bm{n},t),
\end{equation}
where $\mu_{0}$ is the permeability of vacuum and $P(\bm{r}, \bm{n},t)$ is the probability to find a magnetic swimmer located at $\bm{r}$ and pointing along $\bm{n}$ at time $t$.

%Equations (\ref{L1}), (\ref{L2}), and (\ref{bmtor_potential}),
The above equations describe a many-body interacting system with coupled stochastic dynamics of position and orientation under the influence of intrinsic activity, external driving, and thermal noise. To gain insight into the behavior of this complex system, we start by using a number of approximations. By assuming a separation of time scales between the orientation relaxation and the characteristic time scale for positional structure formation, we can formally solve Eq. (\ref{L2}) to find the orientation stationary state $\bm{n}_{\mathrm{st}}(\bm{r})=-\sin\theta(\bm{r}) \;\bm{e}_{r}-\cos\theta(\bm{r}) \;\bm{e}_{z}$ [see Fig.~\ref{sketch}(b) for the definition of $\theta$]. This approximation amounts to the following form for the distribution function $P(\bm{r}, \bm{n})=\rho(\bm{r}) \delta(\bm{n}-\bm{n}_{\mathrm{st}}(\bm{r}))$, where $\rho(\bm{r})$ is the density profile.

\paragraph{Driven swimmer.---}
Ignoring the dipole-dipole interactions, the average stationary state will be determined by the competition between the magnetic torque and the vorticity of the flow, leading to $\sin \theta \simeq v_{\mathrm{f}} k_{\rm B}T r/\left(D_{r}R_{0}^{2}m_{0}B_{\mathrm{ext}}\right)$; i.e. the average orientation of the swimmer will depend on its radial position $r$. The stationary orientation will exist only for sufficiently strong magnetic fields that satisfy the condition $\frac{m_{0}B_{\mathrm{ext}}}{k_{\rm B}T} \geq \frac{v_{\mathrm{f}}}{D_{r}R_{0}}$,
which applies to the experiment reported in Ref. ~\cite{Waisbord2016} (for which $m_{0}B_{\mathrm{ext}}\sim 20 \;k_{\rm B}T$). At the same level of approximation, the translational motion of the swimmer in the radial direction can be described as an Ornstein-Uhlenbeck (OU) process \cite{Waisbord2016}, where the swimmer can be regarded as being constrained in an effective quadratic potential well ${\cal U}_{\rm eff}=\frac{1}{2} \left(\frac{ v_{\mathrm{f}} m_{0}B_{\mathrm{ext}}}{v_{0} R_{0}^{2}}\right) r^{2}$. Hence, magnetotactic bacteria in a Poiseuille flow will exhibit radial focusing under sufficiently large external magnetic field, described by a Gaussian stationary distribution $\rho(r) \sim e^{-r^{2}/(2R^{2})}$, with the characteristic focusing radius $R$ defined as
\begin{equation}\label{R-def}
R^{2}=\frac{v_{0}k_{\rm B}T}{v_{\mathrm{f}}m_{0}B_{\mathrm{ext}}}\; R_{0}^{2}.
\end{equation}
Note that for the cases we are interested in, $R$ is typically much smaller than $R_0$. The time scale for relaxation into this distribution in the radial direction is ${\cal T}_{\rm OU}={ D_{r}R_{0}^{2}m_{0}B_{\mathrm{ext}}}/\left({v_{0}v_{\mathrm{f}}k_{\rm B}T}\right)$.
\begin{figure*}
\begin{center}
\includegraphics[width=2\columnwidth]{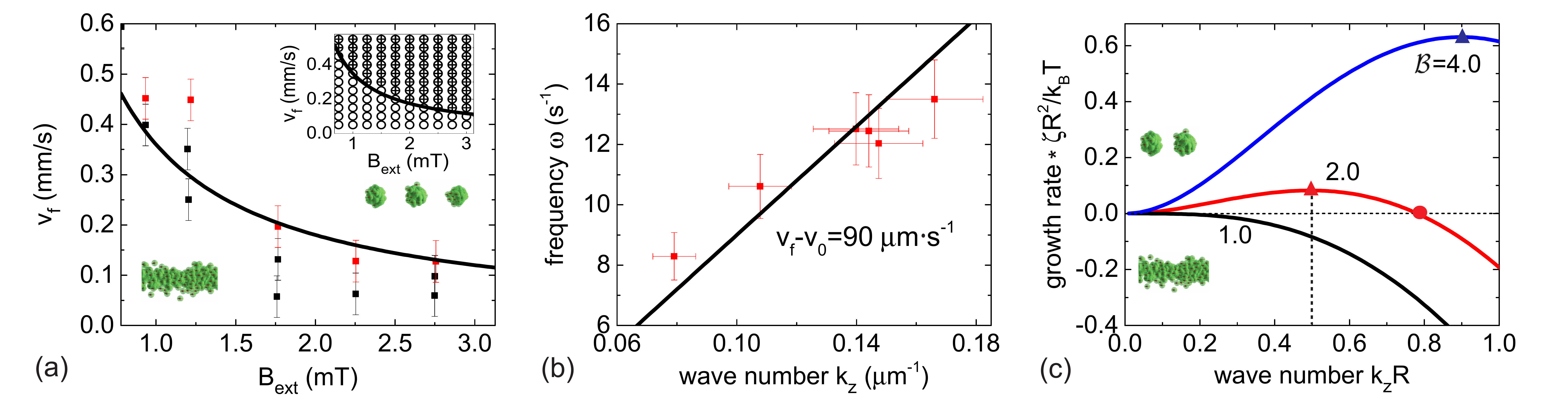}
\caption{(a) Comparison between the prediction of Eq.~(\ref{condition}) for the threshold of clustering (solid line) and the experimental data extracted from Ref.~\cite{Waisbord2016} (where the black/red squares denote the experiments that led to non-clustering/clustering, respectively), in ($v_{\mathrm{f}}, B_{\mathrm{ext}}$) space.
Upper inset: comparison between the theoretical prediction and our Brownian dynamics simulation for the threshold of clustering ($\circ$ for non-clustering and $\oplus$ clustering).
The solid line represents the critical condition for clustering instability given in Eq. (\ref{condition}). Snapshots are adapted from our simulations. (b) Fitting of the dispersion relation $\omega=k_{z}(v_{\mathrm{f}}-v_{0})$ with propagation speed $(v_{\mathrm{f}}-v_{0})=90\;\mathrm{\mu m} \cdot s^{-1}$, corresponding to $B_{\mathrm{ext}}=2.8\;\mathrm{mT}$ and $v_{\mathrm{f}}=180\;\mathrm{\mu m} \cdot \mathrm{s}^{-1}$.
(c) Growth rate versus wave number in systems with $\cal{B}=$$1.0$, $2.0$, and $4.0$ defined in Eq.~(\ref{condition}) ($\blacktriangle$ denotes the maximum growth rate).}
\label{compare}
\end{center}
\end{figure*}

\paragraph{Driven interacting swimmers.---}
We now examine the effect of the magnetic dipole-dipole interaction between the swimmers for a finite (average) density $\rho_{0}$. For time scales longer than ${\cal T}_{\rm OU}$ when the swimmers have reached the stationary state in the radial direction of the channel, it is a good approximation to use the following expression for the density profile of the swimmers
\begin{equation}\label{probability}
\rho(r,z;t) = N  \exp\left({-\frac{r^{2}}{2R^{2}}}\right) c(z;t),
\end{equation}
where the normalization constant is given as $N^{-1}=2\pi R^{2}\;[1-\exp(-R_{0}^{2}/2R^{2})]\simeq 2\pi R^{2}$, and azimuthal symmetry is assumed. This decomposition
%is analogous to the lubrication approximation in thin liquid films, and
is justified provided the characteristic length scales along the channel are considerably larger than the focusing radius. This form allows us to formulate an effective description of the dynamics of the system along the channel.

When the external magnetic field is sufficiently strong, we can invoke the approximation $\bm{n}_{\rm st}\cdot\bm{e}_{z}=-\cos\theta\simeq-1$, which allows us to determine the internal magnetic field ${\bm{B}}_{\mathrm{int}}$ in terms of the density $\rho$ in Fourier space (defined via $\hat{\rho}(\bm{k})=\int d\bm{r}\; e^{-i\bm{k}\cdot\bm{r}} \rho(\bm{r})$ etc) from Eq.~(\ref{bmtor_potential}) as \cite{suppl}
\begin{equation}\label{magneticfield}
\hat{\bm{B}}_{\mathrm{int}}(\bm{k})=-\frac{\mu_{0}m_{0} }{k_{z}^{2}+k_{\perp}^{2}} \left(k_{\perp}^{2}\bm{e}_{z}- k_{z}\bm{k}_{\perp}\right) \hat{\rho}(\bm{k}),
\end{equation}
where $\bm{k}_{\perp}=(k_{x},k_{y},0)$. By inserting Eqs. (\ref{probability}) and (\ref{magneticfield}) in the 3D Fokker-Planck equation for $\rho(r,z)$ and integrating over the radial coordinate, we can obtain an effective 1D Fokker-Planck equation of the slowly varying concentration field $c(z;t)$ as follows \cite{suppl}
\begin{equation}\label{FP3-1}
    \partial_{t}c(z;t)=-\partial_{z} \left[c(z;t)\left(v_{\mathrm{f}}-v_{0}-\frac{1}{\zeta}\;\partial_{z}\left(\frac{\delta\mathfrak{F}}{\delta c}\right)\right)\right],
\end{equation}
where $\mathfrak{F}$ is an effective free energy %Cahn-Hilliard free energy~\cite{Bray2002} %,Speck2014}
given as
\begin{equation}\label{Cahn1}
    \mathfrak{F} =k_{\rm B}T \int dz \; \left(c\ln c-c \right)+\frac{1}{2} \int d z dz' \;{\cal G}(z-z') c(z) c(z'),
\end{equation}
which consists of entropic and energetic contributions. The interaction term involves a long-range kernel defined as
\begin{math}
{\cal G}(z)=- \frac{\mu_{0}m_{0}^{2}}{4\pi R^{2}}\int \frac{dk_{z}}{2\pi}e^{ik_{z} z } g(k_{z}R),
\end{math}
where $g(q)=\left[1+q^{2}\exp(q^{2})\mathrm{Ei}(-q^{2})\right]$ is a monotonically decreasing function, with its maximum value being $g(0)=1$ \cite{suppl}.

\paragraph{Short-time behavior.---}
To examine the dynamics of the interacting system, we first probe the stability of the uniform distribution (that we obtained in the dilute limit) by inserting $c(z;t)=\rho_{0}(\pi R_{0}^{2})+\delta c(z;t)$ in Eq. (\ref{FP3-1}) and solving for the dynamics of $\delta c$ to the linear order. We obtain $\delta \hat{c}(k_{z},t)=\delta \hat{c}(k_{z},0) e^{\lambda(k_z) t}$ where
\begin{equation}\label{phaserelation}
\lambda(k_z)=\Big[\frac{\mu_{0}m_{0}^{2}\rho_{0} R_{0}^{2}}{4 R^{2}} g(k_{z}R)-k_{\rm B}T\Big]\frac{k_{z}^{2}}{\zeta}-ik_{z}\left(v_{\mathrm{f}}-v_{0}\right).
\end{equation}
The real term on the right-hand side of Eq.~(\ref{phaserelation}) determines the growth rate of the wave with given wave number $k_{z}$, and the imaginary term describes a wave propagation with velocity $v_{\mathrm{f}}-v_{0}$ (the sign denotes the direction of propagation). Equation (\ref{phaserelation}) predicts that the system will be unstable to formation of longitudinal density waves, which lead to periodic clustering of the swimmers, provided the following criterion is satisfied
\begin{equation}\label{condition}
{\cal B} \equiv \frac{\mu_{0}\rho_{0} m_{0}^{2}}{4k_{\rm B}T}\frac{m_{0}B_{\mathrm{ext}}}{k_{\rm B}T}\frac{v_{\mathrm{f}}}{v_{0}}\geq 1.
\end{equation}
Therefore, the stability of the magnetic swimmer suspension towards clustering is determined by a subtle balance between the magnetic dipole-dipole interaction and the external magnetic alignment strength versus thermal energy, as well as a competition between the flow and swimming speeds.

We can now compare this result with the experiment of Waisbord \emph{et al.} \cite{Waisbord2016}, which had the following parameters: magnetic moment $m_{0}=1.0\pm0.2\times10^{-16} \;\mathrm{A}\cdot\mathrm{m}^{2}$, self-propulsion speed $v_{0}=100\pm10 \; \mathrm{\mu m }\cdot \mathrm{s}^{-1}$, temperature $T\simeq310\;\mathrm{K}$, number density $\rho_{0}\sim10^{16}\;\mathrm{m}^{-3}$ and width of the (square) channel $50\pm10~\mathrm{\mu m}$. In Fig.~\ref{compare}(a), we show a comparison between the experimental values observed for the threshold of cluster formation, and the solid line (above which clustering occurs) that is a plot of Eq. (\ref{condition}) for the swimmer number density $\rho_{0}=1.6\times10^{16}\; \mathrm{m}^{-3}$ (slightly larger than the experimental value) and other parameters taken as above. Considering the simplifying assumptions in the model, e.g. in the location of the magnetic dipole and the assumption of a circular cross section as well as the absence of hydrodynamic interactions, this can be regarded as a reasonable agreement between theory and experiment. The focusing radius is obtained as $R=0.11\;R_{0}\simeq 2.7\;\mathrm{\mu m}$ in a circular channel of radius $25\;\mathrm{\mu m}$.
We also perform Brownian dynamics simulations based on Eqs. (\ref{L1}) and (\ref{L2}) with $10^{3}$ magnetic swimmers \cite{suppl}, and find the same clustering condition [inset of Fig.~\ref{compare}(a)].
Figure \ref{compare}(b) shows a comparison between the dispersion relation of the propagating wave between theory and experiment, which gives a propagation speed of $v_{\mathrm{f}}-v_{0}=90\;\mathrm{\mu m} \cdot \mathrm{s}^{-1}$, that coincides with the experimental value ($v_{\mathrm{f}}=180\;\mathrm{\mu m} \cdot \mathrm{s}^{-1}$ and $v_{0}=100\pm10 \;\mathrm{\mu m }\cdot \mathrm{s}^{-1}$). While in this experiment the bacteria move downstream, it is also possible to have the opposite case where the net drift will be upstream.

In the clustering region, Eq. (\ref{phaserelation}) predicts a band of exponentially growing modes, as can be seen in Fig.~\ref{compare}(c), which shows the growth rate for different values of ${\cal B}$. We can estimate the time scale for triggering the most probable wave pattern (with maximum growth rate) using the inverse of the growth rate.
%As shown in Fig.~\ref{compare}(c), the triggering time decreases with increasing ${\cal B}$.
For the experimental parameters $B_{\mathrm{ext}}=2.1\; \mathrm{mT}$ and $v_{\mathrm{f}}=300\; \mathrm{\mu m }\cdot \mathrm{s}^{-1}$~\cite{Waisbord2016}, which correspond to $\mathcal{B}=1.7$, the obtained estimate is about $10^{2}\; \mathrm{s}$; this is one order of magnitude larger than the clustering time in the experiment ($\sim10\; \mathrm{s}$)~\cite{Waisbord2016}. In our numerical simulations, the time scale for clustering is about $10\; \mathrm{s}$, which coincides with the experimental value \cite{suppl}. This may indicate that the nonlinear regime of the dynamics determines the overall time scale of cluster formation. However, one should also note the sensitivity of this time scale to the experimental parameters. For example, if we change $m_{0}$ from $10^{-16} \;\mathrm{A}\cdot\mathrm{m}^{2}$ to $1.2\times 10^{-16} \;\mathrm{A}\cdot\mathrm{m}^{2}$, which is within the reported range (corresponding to $\mathcal{B}=3.0$), the theoretical estimate will also become $\sim10\; \mathrm{s}$.

We can determine the fastest growing mode analytically in the regime ${\cal B} \gtrsim 1$ by using an approximation of the function $g(k_{z}R)$ in the limit $k_{z}R\ll 1$, in the form of $g(k_{z}R) \simeq %1+\left[2\ln(k_{z}R)+\gamma\right] (k_{z}R)^{2}\simeq
1-g_2 k_{z}^{2}R^{2}$, where $g_2$ is a number of order unity \cite{suppl}. This gives an approximate expression for the growth rate as
\begin{math}
\left[\rho_{0} \pi R_{0}^{2}(\mathcal{K}-\mathcal{M}k_{z}^{2})-k_{\rm B}T\right] k_{z}^{2}/\zeta
\end{math}
where $\mathcal{K}=\frac{\mu_{0}m_{0}^{2}}{4\pi R^{2}}$ and $\mathcal{M}=\frac{g_2\mu_{0}m_{0}^{2}}{4\pi}$. Using this approximation, the maximum growth rate can be found as $k_{z}^{\mathrm{max}}\simeq \frac{1}{R}\sqrt{\frac{{\cal B}-1}{2 g_2 {\cal B}}}$. Hence, the most prominent wavelength for the instability will be determined by the focusing radius and the distance from the threshold of instability. 

\paragraph{Long-time behavior.---}
To determine the long-time behavior of the system, we need to study the full nonlinear dynamics described by Eqs. (\ref{FP3-1}) and (\ref{Cahn1}). Using the above expansion for $g(k_{z}R)$, we can approximate the free energy as
\begin{equation}\label{Cahn2}
\mathfrak{F} \simeq \int dz\left[k_{\rm B}T(c\ln c-c)-\frac{\mathcal{K}}{2}\;c^{2}+\frac{\mathcal{M}}{2}(\partial_{z}c)^{2}\right],
\end{equation}
which resembles a Cahn-Hilliard free energy \cite{Bray2002}. This shows that the effective dynamics can lead to a phase separation in the moving frame with velocity $v_{\mathrm{f}}-v_{0}$, resulting from a competition between entropic effects and attractive dipole-dipole interaction. The onset of the tendency towards phase separation coincides with the threshold of instability [given in Eq. (\ref{condition})], as can be seen from the expansion of Eq. (\ref{Cahn2}) in powers of density fluctuations:  $\mathfrak{F} \simeq k_{\rm B}T \rho_0 \pi R_0^2 \int dz [\frac{1}{2}(1-{\mathcal B}) (\rho_0 \pi R_0^2)^{-2} \delta c^2-\frac{1}{6} (\rho_0 \pi R_0^2)^{-3} \delta c^3+\frac{1}{12} (\rho_0 \pi R_0^2)^{-4} \delta c^4+\cdots+\frac{\mathcal{M}}{2}(\partial_{z}\delta c)^{2}]$, where we have ignored a constant term, and a term linear in $\delta c$ that will lead to a shift in the Lagrange multiplier than enforces number conservation. The dense phase will form a single domain that accumulates most of the swimmers and moves with the speed of $v_{\mathrm{f}}-v_{0}$ in coexistence with the dilute phase. Since in the long-time limit the system is effectively one dimensional, this macroscopic demixing will be susceptible to noise~\cite{Bray2002}, which means that the front will intermittently dissociate due to fluctuations triggered by noise and then re-form.

\paragraph{Hydrodynamic interactions.---} %Pooley2007,Yeomans2014,
It is important to examine whether the above discussions are qualitatively modified if we incorporate hydrodynamic interaction between the swimmers, which can be repulsive or attractive depending on swimmer types~\cite{AditiSimha2002,Saintillan2008,Lauga2017,Elgeti2015}. To account for this effect, we consider an active stress in the fluid (arising from the hydrodynamic activity of the swimmers) in the form of $\bm{\Sigma}_{a}= S \rho \left(\bm{e}_{z}\bm{e}_{z}-\frac{\bm{1}}{3}\right)$,
where $S$ denotes the strength of the force dipoles exerted by the swimmers on the fluid (stresslet)~\cite{Batchelor1971} ($S<0$ for pushers and $S>0$ for pullers). %Nash2010,Karimi2013,
The presence of a confining wall affects hydrodynamic interactions~\cite{Li2014,Wioland2016,Singh2017}. For two stokeslets in a cylindrical channel, this effect is negligible if their distance $d$ is considerably smaller than the radius $R_0$ (e.g. $d/R_{0}\lesssim 10^{-1}$), while the hydrodynamic interaction is effectively screened for relatively large distances ($d/R_{0}\geq 1$)~\cite{Liron1978,happel}. Here, the swimmers are densely packed along the axis of the channel as $R \ll R_{0}$, and therefore the effect of the wall is negligible for distances smaller than $R_0$ (see below).
Moreover, the hydrodynamic effect due to the drag caused by the magnetic dipolar forces can also be ignored as it is weaker than the magnetic dipole-dipole interaction itself.
By solving the Stokes equation for an incompressible fluid with this active stress, we obtain a fluid flow ${\bm u}({\bm r})$ that will be superimposed with $\bm{V}_{\mathrm{f}}$ in Eqs. (\ref{L1}) and (\ref{L2}). In Fourier space, we obtain $\hat{\bm u}({\bm k})=\frac{i S}{\eta (k_{z}^{2}+k_{\perp}^{2})^2} \left(k_{\perp}^{2} k_z \bm{e}_{z}- k_{z}^2 \bm{k}_{\perp}\right) \hat{\rho}(\bm{k})$ \cite{suppl}. By incorporating the longitudinal component of the flow into the calculations, Eq. (\ref{phaserelation}) will be modified as follows
\begin{eqnarray}\label{condition_2}
\lambda(k_z)&=&\left[\frac{\mu_{0}m_{0}^{2}\rho_{0} R_{0}^{2}}{4 R^{2}} g(k_{z}R)+\frac{S\rho_{0} R_{0}^{2}\zeta}{4\eta}h(k_{z}R)-k_{\rm B}T\right]\frac{k_{z}^{2}}{\zeta}\nonumber\\
&&-ik_{z}\left(v_{\mathrm{f}}-v_{0}\right),
\end{eqnarray}
where $h(q)=-1-\left(1+q^{2}\right)e^{q^{2}}\mathrm{Ei}(-q^{2})$ \cite{suppl}.
As we ignore the effect of the confining wall on the hydrodynamic interaction between swimmers, this equation is only valid if the characteristic wave length is smaller than the size of the channel.

Since both $g(q)$ and $h(q)$ are positive, Eq. (\ref{condition_2}) indicates that for aligned pushers ($S<0$) hydrodynamic interactions resist clustering, whereas for aligned pullers ($S>0$) they favor clustering. The growth rate is plotted for various values of $\mathcal{H}={S\rho_{0} R_{0}^{2}\zeta}/{(4\eta k_{\rm B}T)}$ in the Supplemental Material, where a diagram summarizing the different possibilities is also provided \cite{suppl}.
%The condition for clustering incorporating the effect of the hydrodynamic interactions between swimmers then approximately becomes ${\cal B}+\mathcal{H} h_0\geq  1$, where $h_0$ is the value of the function $h(q)$ at the characteristic value of the fastest growing mode; this is typically $q \sim 1$, which yields as estimate of $h_0 \sim 10^{-2}-10^{-1}$ \cite{suppl}.
%[see Eq. (\ref{condition})].
If the effect of the hydrodynamic interaction in Eq. (\ref{condition_2}) is comparable with that of thermal fluctuations, i.e. $\left|S\right| \rho_{0} R_{0}^{2}\zeta/(4\eta) \sim k_{\rm B}T$, then the critical line for clustering  in Fig.~\ref{compare}(a) is shifted in the ($v_{\mathrm{f}}$, $B_{\mathrm{ext}}$) plane. If the effect of the hydrodynamic interaction is much stronger than that of thermal fluctuations, then for pushers the clustering will be determined by the competition between the magnetic dipole-dipole interaction and the hydrodynamic interaction only. This may apply to synthetic magnetic microswimmers~\cite{Peyer2013,Erb2016}, which can be designed as pushers and can carry large magnetic moments, but is unlikely to apply to the magnetotactic bacteria (\emph{Magnetococcus Marinus} MC1) used in Ref.~\cite{Waisbord2016} since their magnetic dipole moment cannot be strong enough to allow the dipole-dipole interaction to compete with the hydrodynamic interaction. Pullers, on other hand, always form clusters, regardless of the strength of the magnetic dipole-dipole interaction. We also note that in the experiment of Ref.~\cite{Waisbord2016} the bacteria cannot form clusters until they are focused above a certain threshold, which provides a sufficiently high density where the magnetic dipole-dipole interaction is comparable with the thermal energy. In light of the above arguments, we can thus surmise that in this experiment the hydrodynamic interaction between the bacteria is most likely either negligible or at most comparable with thermal fluctuations, and thus, also with the magnetic dipole-dipole interaction. We note that the value of $S$ in our simple description corresponds to the time-averaged value of the stresslet moment, which could be significantly smaller than instantaneous values, as has been shown for the case of \emph{Chlamydomonas}, due to oscillations between puller and pusher behavior in a stroke cycle~\cite{Klindt2015}.

In conclusion, we illustrate how the combination of external magnetic alignment, magnetic dipole-dipole interaction, thermal fluctuations, self-propulsion, and external shear flow work together to determine the collective behavior of magnetic swimmers. We have shown that magnetic interactions can explain the experimental observation of clustering. We believe our results exemplify the tunability and the diverse range of behavior that can be achieved in magnetic active matter.

\begin{acknowledgements}
We would like to acknowledge helpful discussions with Julia Yeomans. This work has received funding from the Horizon 2020 research and innovation programme of the EU under grant agreement No. 665440, and the COST Action MP1305 Flowing Matter.
\end{acknowledgements}

%\end{document}

\end{document}